# Analyzing Popularity of Software Testing Careers in Canada

**Pradeep Kashinath Waychal, Western Michigan University,**

**Luiz Fernando Capretz, Western University, London, Ontario, Canada**

**Sachin Narendra Pardeshi, R.C. Patel Institute of Technology, India**

**Introduction**

As software systems are becoming more pervasive, they are also becoming susceptible to failures, resulting in potentially lethal combinations. There have been catastrophic failures such as Ariane 5[2], Therac-25[3], and the UK e-borders project[4], which led to the loss of life and capital. Many similar incidents are happening all over the world[5]. A Micro Focus[6] report points out that the effects of software failures are influencing discussions in boardrooms and even brand names. Even though the software industry has been using advanced technologies and processes for development activities, software failures have not decreased[7].

Software testing is critical to prevent software failures. Therefore, research has been carried out in testing but that is largely limited to the process[8,9] and technology[10,11] dimensions and has not sufficiently addressed the human dimension. Even though there are reports about inadequacies of testing professionals and their skills [6], only a few studies have tackled the problem[12]. Therefore, we decided to explore the human dimension. We started with the basic problem that plagues the testing profession, the shortage of talent[6], by asking why do students and professionals are reluctant to consider testing careers, what can be done about that, and is the problem specific to locales or spread across the globe? This paper focusses on these questions.

We have studied unpopularity of testing careers among students and professionals earlier in the Indian context[13]. The study has pointed out the need to investigate the problem in other geographies to develop better understanding of the problem, given the criticality of the situation. Towards that, we chose Canada as it is significantly different than India on some key parameters such as the networked readiness index, per capita GDP rank, contribution of IT to the national GDP, and unemployment rate that could impact career choices.

Towards that end, we carried out a survey among senior students and alumni of a reputable Canadian software engineering program, which was one of the first to be accredited by the Canadian Engineering Accreditation Board (CEAB). We asked the students and alumni to list the PROs and CONs of a testing career and if they would choose that career and the reasons thereof. After analyzing the reasons, we are proposing solutions to bring in changes that would attract more individuals to testing careers. The next section covers the research design process that includes discussion, comparison with Indian students and with working professionals[13].

**Research Design**

Our study analyzed the views of software engineering students and alumni (professionals) about testing careers. We asked a sample of students if they would like to choose testing careers and what they felt were the PROs and CONs of the testing careers. We compared PROs and CONs of students with those provided by professionals to know if the students have a proper understanding of industry and to propose possible solutions. The overall research design is outlined in Figure 1. Since we wanted to compare perspectives of testers in various geographies, we followed the same process and questionnaire that we followed in the Indian study[13].

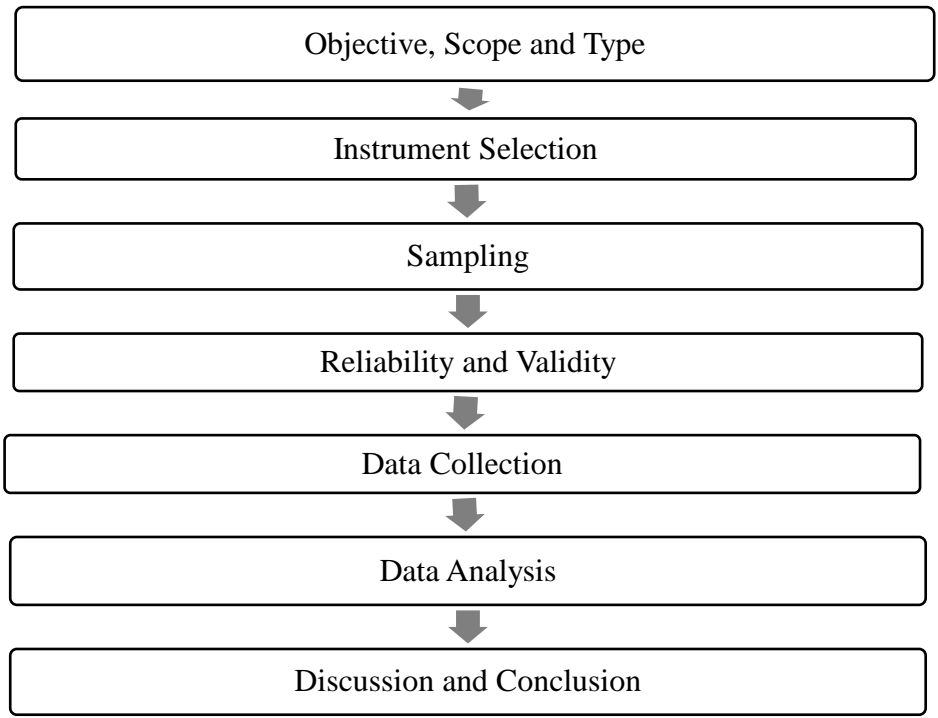

Figure 1: Research design

**Objective, Scope and Type**

Very few bright individuals voluntarily choose testing careers, which robs the industry of good testing and delivery of quality products. To change this situation, it is necessary to analyze the reasons for such apathy towards testing careers. Our study analyzed the reasons for not choosing testing careers by Canadian software engineering students and professionals.

We asked senior students of a software engineering program, if they would choose testing careers, and what they saw as PROs and CONs of these careers. The research is descriptive, diagnostic, cross-sectional, and mixed. Descriptive research describes the characteristics of a population being studied and does not explore the reasons for those characteristics. Diagnostic research studies determine the frequency with which something occurs or its association with something else. We did not study the event over time but at a cross-section, making the study cross-sectional. We used a qualitative method by asking open-ended responses to the PROs and CONs of testing careers and quantitative method by asking categorical answers about choosing testing careers, making the study a mixed one.

**Instrument Selection**

We asked students for the probability that they would choose testing careers by offering multiple choices: "Certainly Yes," "Yes," "Maybe," "No," and "Certainly Not." We have not found adequate number of prior studies in this area. Therefore, we asked our respondents to provide open-ended but prioritized list of PROs and CONs, and open-ended rationale in support of their decisions.

**Sampling**

Our sample consisted of 85 senior undergraduate students and 20 alumni (software professionals) of a software engineering program from a Canadian university. We decided to involve alumni, to understand professionals' perspective on testing, due to their easy accessibility and availability to the university.

**Reliability and Validity**

The characteristics of a qualitative study are conceptualized as trustworthiness, rigor, and quality[14]. Lincoln and Guba[15] believe that in the case of qualitative studies validity implies reliability and suggest a demonstration of only validity. Creswell and Miller[16] have observed that qualitative researchers employ member checking, triangulation, peer reviews, thick description, and external audits to demonstrate validity. We asked the respondents to list PROs and CONs of testing careers and the probability that they would choose a testing career along with their rationale. We triangulated rationales and PROs-CONs to find virtually no divergence between them and thickly described the survey responses.

**Data Collection -** *Students*

We explained the background of our study to students in their class sessions and sought their responses on their desire to take up testing careers (Table 1) and on PROs and CONs thereof. We manually tagged all the responses and then iteratively coded them until no further code changes (merging or demerging) were possible. Table 2 provides frequencies of various PROs and Table 3 of various CONs.

Table 1: Chances of taking up testing career

| Response | Number | Percentage |
|---|---|---|
| Certainly Yes | 2 | 2 |
| Yes | 6 | 7 |
| Maybe | 28 | 33 |
| No | 23 | 27 |
| Certainly not | 26 | 31 |
| **Total** | **85** | **100** |

Table 2: Frequencies of PROs of students

| PRO -> | Important Job | Easy Job | More Jobs | Learning Opportunities | More Money | Think-ing Job | Fun to break things |
|---|---|---|---|---|---|---|---|
| Total | 36 | 36 | 35 | 19 | 16 | 14 | 13 |
| % | 21% | 21% | 20% | 11% | 9% | 8% | 7% |

Table 3: Frequencies of CONs of students

| CON -> | Tedious | Less Creativity | 2nd class citizen | Miss development | Finding mistakes of others | Complex-ity/ Stressful |
|---|---|---|---|---|---|---|
| Total | 60 | 40 | 33 | 30 | 12 | 11 |
| % | 30% | 20% | 17% | 15% | 6% | 6% |

**Data Collection -** *Professionals*

We repeated the same exercise with 20 professionals. The exercise – explanation of the background and seeking responses - happened over email.

We asked them the chances of their taking up or continuing with testing careers and tabulated their responses in Table 4. Table 5 and Table 6 provide frequencies of various PROs and CONs, respectively.

Table 4: Chances of taking up testing career by working professionals

| Response | Number |
|---|---|
| Certainly Yes | 3 |
| Yes | 2 |
| Maybe | 6 |
| No | 6 |
| Certainly not | 3 |
| **Total** | **20** |

Table 5: Frequencies of PROs of working professionals

| PRO -> | Learning Opportunities | Important Job | Easy Job | More Jobs /Job Security | Challenging / Thinking Job | More Money |
|---|---|---|---|---|---|---|
| Total | 19 | 9 | 9 | 7 | 4 | 3 |
| % | 34 | 16 | 16 | 13 | 7 | 5 |

Table 6: Frequencies of CONs of working professional

| CON -> | 2nd class citizen | Miss dev / limited learning opportunities | Tedious (Repetitive work) | Complexity / challenging | Less Money | Less creativity /not challenging |
|---|---|---|---|---|---|---|
| Total | 27 | 7 | 6 | 6 | 6 | 4 |
| % | 46 | 12 | 10 | 10 | 10 | 7 |

We did compute weighted frequencies of PROs and CONs by assigning weights of 5,3, and 1.5 to the 1st, 2nd and 3rd entries, respectively, but did not find sufficient differences between simple and weighted frequencies, and therefore decided to include only simple frequencies. We excluded the PROs and CONs, whose frequencies were less than 5%.

**Data Analysis**

The analysis of responses resulted in the following categories of PROs and CONs as described in Tables 7 and 8.

Table 7: Explanation of PROs along with sample statements

| PRO | Sample statements (Verbatim) |
|---|---|
| • Learning opportunities – Testers can learn different products, technologies, techniques, and languages as well as domains such as retail, financial. They can also develop softer skills, due to more (difficult) interactions with developers and customers. Testing activities provide full background of project scope, architecture, and integration strategy in a short period of time and span all project stages. Further, testing requires focusing on details and is a growing field. | • *Get to understand 'ins' and 'outs' of how the system work.*<br>• *Testers have a wider view of the system since they have to work on all the phases of the software life cycle*<br>• *Learn broad knowledge in different applications*<br>• *Improve your communication and technical skills* |
| • Important jobs – Testers are accountable and responsible for the product quality. In that sense testing is an important part of software life cycle. | • *QA is very important role in software development. They focus on finding bugs which (is) different than developers.*<br>• *The opportunity to be involved in producing high quality software* |
| • Easy jobs – This refers to the jobs having well defined and easy processes. | • *Clearly defined objectives and metrics*<br>• *Structured work schedule* |
| • More jobs / Secure jobs / Stable jobs – This states that more testing jobs are available and due to the higher | • *A lot of QA jobs out there.*<br>• *Stable job* |
| demands and lower supplies of testers, the jobs are secure and stable. | |
| • Thinking and creative / Challenging job – This encompasses views about testing such as being challenging, creative, innovative, and requiring logical and analytical thinking. | • *Challenging due to many error possibilities.* <br> • *Improving critical thinking – it's often harder to find what others have missed* |
| • More money – Testing jobs come with good salary packages. | • *Financially rewarding.* <br> • *You get to work in a very lucrative industry.* |
| • Career growth – Some professionals think that testing has better growth | • *It's a means to an end: learn on the "shop floor" and move up to roles such as Solutions Architect, Director - Project Management, Chief Technology Officer, or Entrepreneur.* <br> • *Wide career path and long-term growth* |
| • Fun to break things and finding mistakes of others | • *It's satisfactory to break code and find bugs* |

Table 8: Explanation of CONs along with sample statements

| CON | Sample statements (Verbatim) |
|---|---|
| • Tedious – This refers to the repetitive nature of testing and respondents have also used words such as monotonous and boring. | • *Testing is repetitive work requiring loads of screen time. This is the "digital equivalent" of working as a labourer on a manufacturing assembly line; physically exhausting and mentally boring* <br> • *Repetitive work, some people do not like* |
| • Less creative, not challenging | • *Could be less and less creative* <br> • *There are hardly any challenges* |
| • Second-class citizen – This is a major factor and is commonly voiced by respondents and includes testers not being involved in decision making, and being blamed for poor | • *Second-Tier Professional – testers are typically regarded as second-class citizens within the organization. They have almost "no-say" on the architecture and design* |
| quality, while developers are rewarded for good quality. It also includes the lack of support from management resulting in unrealistic schedules, a scarcity of resources, and the struggle for recognition | *of a system. They are always at the rear end of the development cycle, meaning challenged with very little remaining time to ship the product. The work is also tedious and repetitive. (People tease that monkeys can do this kind of work!) The pressure is high due to time constraint. It is typical to see test teams working overtime over multiple weekends prior to shipping of the system, performing regression tests over and over again, with multiple last-minute bug fixes from the development team. There will always be heated debates on whether defects are qualified or not – whether there are problems with setting up the test environment, whether there are problems with testers understanding the functionality of the system, etc. It is not surprising to arrange overnight stress test (hopefully automated, but with tester on call) to qualify the system for shipment first thing Monday morning in order to meet the deadline.*<br><br>• *If software fails, testers are more responsible than developers.* |
| • Miss development / No coding – This relates to testers not developing code or software. | • *Never get to design software, must follow someone else's code.*<br>• *Creating software can be more exciting than testing software* |
| • Complexity / stressful / frustrating – This set covers testers facing complex situations such as different versions of software, inadequate infrastructural support, platform incompatibilities, defects not getting reproduced, and not being allowed sufficient time, but being held responsible for product quality. This also includes the fact that testers need to look at business and | • *Complexity of writing stubs*<br>• *Unexpected events may happen anytime rendering the performed tasks useless.*<br>• *Difficult to find errors and time consuming.*<br>• *It requires extensive amount of documentation.* |
| technology artefacts and understand many abstractions. The lack of clarity around requirements also adds to the difficulties. | |
| • Less monetary benefits – Some testers believe that testers' jobs do not have good monetary benefits. | • *At least the jobs that I've seen have all been lower paid than the equivalent dev jobs*<br>• *Less income compared to developer.* |
| • Finding the mistakes of others – It is not easy to find out mistakes in others' work and present them. | • *Being seen as evaluating the work of peers, may lead to workplace dissonance and lack of credibility, no matter one's competence.*<br>• *Sometimes team members hate you professionally due to found bugs* |

While two percent students chose the "Certainly Yes" option, seven percent students chose the "Yes" option. Thirty-one percent students vehemently (by selecting "Certainly Not" option) refused to choose the testing career. Twenty-seven percent students would not like to go for testing careers, and thirty-three percent students were unsure (answered "May be") of their plans.

Some students made ambivalent statements in their PROs and CONs such as, "Ability to think increases" as a PRO and "Does not help for innovation" as a CON; "No Coding" as a PRO and "Missing development as a CON; "Interesting Field" as a PRO and "Boring Life" as a CON. Perhaps, they were looking at the situation from different perspectives.

The professionals also were not inclined to join or continue in testing careers. While 45% chose "Certainly Not" or "No" options, only 25% chose "Certainly Yes" or "Yes" options, and 30% were ambivalent.

**Discussion**

It is evident that the testing profession is far from being popular. In case of students, less than 10% were thinking of taking up testing careers. While 33% of the students were ambivalent, 58% showed a disinclination to join the testing profession, 31% of them responding with the "Certainly Not" option. The professionals also were not so much inclined to join or continue in testing careers. While 45% chose "Certainly Not" or "No" options, only 25% chose "Certainly Yes" or "Yes" options, and 30% were ambivalent.

It seems that the students are aware of the PROs of testing careers. First four PROs of the students and the professionals are the same. One difference is the "learning opportunity" is the topmost PRO for the professionals and the fourth for the students. A related difference is students think of learning of tools, product architecture, and languages than of business domain. While developers are far away from the business customers and their problems, testers enjoy their proximity, can learn immensely from them, and perhaps graduate easily into business analyst roles. The students do appreciate the importance of testing activities and are aware that testers are responsible and accountable for product quality. One way to increase these numbers would be to apprise students of the complete product life cycle through real-life projects and exposure to industry processes. Twenty-one percent PROs recognized testing jobs to be easier, and 20 percent that testing has more jobs. The testing jobs were also seen as offering more money (9%) and challenging / thinking jobs (8%). In case of the professionals, "learning opportunities" (34%) was followed by "important job" (16%), "easy job" (16%), "more jobs" (13%), "challenging / thinking jobs" (7%), and "more money" (5%). Overall, the Canadian students and professional have reasonably similar understanding of the positive features of testing jobs.

Many students believe that testing jobs are tedious (30%), lack creative challenges (20%), and, therefore, rob testers of professional development opportunities. Students don't seem to include test automation activity, which is a software development activity that uses scripting languages and environments like development. The students are aware that the profession is relegated to second-class citizenship (17%). They have cited reasons such as not being involved in the decision-making process, not getting credited for good quality products but getting blamed for bad quality products, not having competitive growth paths, and exerting schedule pressure on testers to compensate for developers' overruns. Some of these problems are relatively easy to fix and they must be fixed. Some students also believe that they will miss development (15%). Interestingly, the professionals also have the same three CONs at the top. The difference is the order. In case of professionals "second-class citizen" (46%) is followed by "miss development" (12%), and "tediousness" (10%). If students are exposed to this reality of the CONs, many more may get distracted from the testing profession. Interestingly, a few students, who were certain about not taking up testing careers had provided reasons such as xxx taught the course and it was a terrible experience. This reinforces the role of faculty in students' career choices.

We also compared the PROs and CONs of Canadian students and professionals with Indian students based on our earlier study[13] (Figure 2). While Indian students regard testing as thinking jobs that provides more learning opportunities, Canadian students regard testing as an area with easy and more jobs. The Indian software sector has plenty of jobs and, therefore, perhaps Indian students do not worry about jobs and do not see that as a PRO. It is also possible that Indian students, unlike Canadian students, learnt testing as a thinking job that offers many learning opportunities, or it is possible that testing jobs in Indian industry are indeed different. In the case of CONs, the Canadian and Indian students seem to converge. The top three CONs are tediousness, less creativity and being second-class citizen. While Indian students are more worried about the second-class citizenship, the Canadian students are worried about tediousness of the job. This again points out to possible differences in testing jobs in the two countries.

The Indian professionals believe that the testing jobs are indeed thinking jobs and allow many learning opportunities. For Canadian professionals, the learning opportunities, importance and ease of jobs appear to be appealing. On the CONs side, again, the Indian and Canadian professionals seem to converge. A difference is Indian professionals see the second-class citizenship issue to be of very grave nature. The Canadian professionals voice that issue but do not seem to believe it to be so grave. The industry leaders, certainly, must work on these aspects, or else they will not get good testing professionals, which will impact software quality and business prospects. It seems that both Canadian and Indian students have reasonable understanding of the industry scenarios resulting in notable overlap in case PROs and CONs with their professional counterparts.

As discussed earlier, we chose Canada for our second leg of study due to significant differences in some key parameters between India and Canada (Table 9). Despite those differences, the testing career is almost equally unpopular among students of both the countries. The reasons for the unpopularity, perhaps, differ due to those parameters. While Canada is much better on "networked readiness" and "total GDP", India is better on "contribution of IT to the national GDP" and "unemployment rate". The Canadian students, therefore, speak strongly of availability of jobs as compared to their Indian counterparts. India's significantly higher contribution of IT to the national GDP despite lower networked readiness and national GDP indicates that India is exporting its IT services. Indian testers, therefore, may be getting chances to work on more challenging testing assignments, giving rise to significant differences in their choosing "thinking jobs" and "learning opportunities" PROs as against their Canadian counterparts. Perhaps due to the same reasons, Indian students are more sensitive to the "second-class citizen" status meted out to them and speak about that more seriously. These significant differences influence importance of other common PROs such as "important job", "easy jobs", and common CONs such as "tedious" and "miss development", i.e., the Indian students don't think so much about them in the face of more seriously perceived PROs and CONs.

Table 9: Key parameters that may affect career choices of students and professionals

| Parameter | Canada | India |
|---|---|---|
| *The Networked Readiness Index[1] | 10 | 91 |
| Per Capita GDP Rank[2] | 24 | 124 |
| Unemployment rate[3] | 5.9% | 3.4% |
| Contribution of IT to the national GDP | 4.4%[4] | 9.3%[5] |

[1]http://online.wsj.com/public/resources/documents/GITR2016.pdf
[2]https://knoema.com/sijweyg/world-gdp-per-capita-ranking-2017-data-and-charts-forecast
[3]https://en.wikipedia.org/wiki/List_of_countries_by_unemployment_rate
[4]https://www.ic.gc.ca/eic/site/ict-tic.nsf/eng/h_it07229.html
[5]granthaalayah.com/Articles/Vol5Iss6/01_IJRG17_A06_327.pdf
*Measures the drivers of the ICT revolution

| Frequencies of PROs – Canadian and Indian Students | Frequencies of CONs – Canadian and Indian Students |
|---|---|

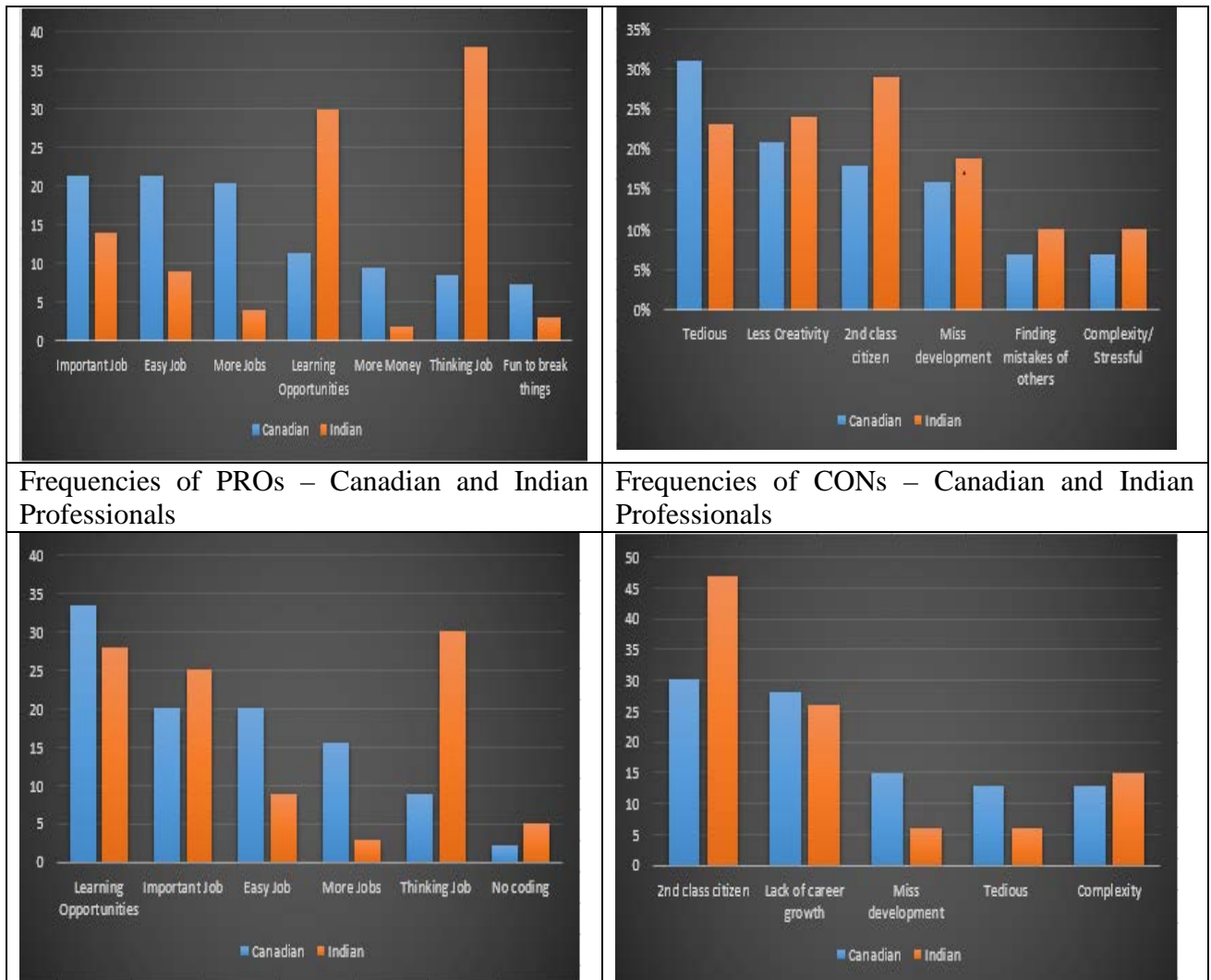

Figure 2: Comparing PROs and CONs of Indian and Canadian students and professionals

It is worthwhile to compare our findings with that of Deak et al.[17] based on their study of 161 Norwegian students from three different classes. They found 58% students not being interested, 17% being ambivalent, and only 25% being interested in the testing jobs. Their top three CONs were "boring", "rather writing code", "not creative"; which we have categorized as "tedious", "miss development" and "less creative". Their "status" and "unrewarding", which were at the next rungs, can be mapped to our "second-class citizen". The top PROs were testing being "interesting" and "important". We can map the "interesting" to "learning opportunities" from our study. Deak et al. also have carried out a qualitative study of close to 40 testing professionals[18] and found out "lack of influence and recognition" and "being unhappy with the management" as the topmost issues. Those two issues, along with their another issue of "time pressure", can be mapped to our "second-class citizen" CON. Besides they also talk about "technical issues", "boredom", and "poor relationship with developers", which can be mapped to "complexity", "tediousness", and "finding mistake of others", respectively. On the PROs side their study found out "Enjoy challenges", "Focus on improving the quality", and "Variety of work", which can be mapped to "challenging jobs", "important jobs", and "learning opportunities". Thus, the Norwegian study's findings have some similarities with our findings.

**Conclusions**

Testing appears to be a neglected area in the software industry. There are not enough testing specialists and test schedules are squeezed as development overruns occur and delivery milestones are considered non-negotiable. Many times, testing is perceived as a nuisance that is sandwiched between development and deployment, when it is a critical activity that needs to be performed in parallel to design and development activities, as advocated by V&V model.

In our study at a reputed Canadian university, we found that very few senior software engineering students and professionals have testing careers on their minds despite various challenges and learning requirements associated with testing jobs. The students have a reasonably good understanding – in terms of the PROs and CONs – of testing careers and still do not want to take them up. In fact, developing a better understanding may dissuade them away from testing. That perhaps explains why the software industry has been facing a shortage of software testers and, as a result, has been facing quality problems.

We also compared our findings from the Canadian university with that from an Indian college. While Indian students regard testing as a "thinking job" that provides "more learning opportunities"; Canadian students regard testing as an area with "easy and more jobs". In the case of CONs, the Canadian and Indian students seem to converge with the top three CONs as "tediousness", "less creativity" and "being second-class citizens". A notable difference is that for the Indian students. the "second-class citizen" is the top most CON. The Indian professionals believe that the testing jobs are indeed "thinking jobs" and allow "many learning opportunities". For Canadian professionals, the "learning opportunities", "importance and ease of jobs" appear to be appealing. On the CONs side, again, the Indian and Canadian professionals seem to converge. A difference is Indian professionals see the "second-class citizenship" issue to be of greater importance. Deak et al.'s [17,18] findings based on their Norwegian study appear to be broadly in line with our findings.

Software testers should be treated with respect and viewed as essential to product success. To reinforce this view, potential software testers should be offered a varied and rewarding career, and better growth opportunities. To identify career paths, industry leaders need to define various roles of a software test practitioner and define a varied and rewarding career paths with potential lateral transfers to other paths, establish appropriate training for associated competencies, determine relevant certification opportunities, and recognize outstanding software testing engineers. Achieving these goals should in turn lead to uniform, efficient, and effective software testing practices, resulting in shortened product development and maintenance cycles, and more reliable products.

Further, the industry and colleges (especially faculty members who teach the testing courses) need to create awareness about these steps among college students so that more software engineering students begin to choose testing careers. It will be worthwhile to study interactions of gender and academic performance with the testing career choices, and PROs and CONs thereof, for students and professionals.

The study was carried out in one college in Canada and its findings are compared with a college in India. Studies in more colleges is required to develop acceptable national views. It also may help to study this phenomenon in more countries and develop global perspectives on the issue. However, the study certainly offers useful insights and helps educators and industry leaders to come up with an action plan to change the outlook towards testers in industry and in computer science and software engineering programs, and put the software testing profession under a new light. That could increase the number of software engineers deciding on testing as a career of their choice, could increase the quality of software testing, and improve the overall productivity, and turnaround time of software development activity.

**Acknowledgments**